# QFL: Data-Driven Feedback Loop to Manage Quality in Agile Development

Lidia López
Universitat Politècnica de Catalunya Barcelona, Spain
llopez@essi.upc.edu

Alessandra Bagnato
Softeam
Paris, France
alessandra.bagnato@softeam.fr

Antonin Ahbervé
Softeam
Paris, France
antonin.ahberve@softeam.fr

Xavier Franch
Universitat Politècnica de Catalunya Barcelona, Spain
franch@essi.upc.edu

*Abstract— **Background**: Quality requirements (QRs) describe desired system qualities, playing an important role in the success of software projects. In the context of agile software development (ASD), where the main objective is the fast delivery of functionalities, QRs are often ill-defined and not well addressed during the development process. Software analytics tools help to control quality though the measurement of quality-related software aspects to support decision-makers in the process of QR management. **Aim**: The goal of this research is to explore the benefits of integrating a concrete software analytics tool, Q-Rapids Tool, to assess software quality and support QR management processes. **Method**: In the context of a technology transfer project, the Softeam company has integrated Q-Rapids Tool in their development process. We conducted a series of workshops involving Softeam members working in the Modelio product development. **Results**: We present the Quality Feedback Loop (QFL) process to be integrated in software development processes to control the complete QR life-cycle, from elicitation to validation. As a result of the implementation of QFL in Softeam, Modelio's team members highlight the benefits of integrating a data analytics tool with their project planning tool and the fact that project managers can control the whole process making the final decisions. **Conclusions**: Practitioners can benefit from the integration of software analytics tools as part of their software development toolchain to control software quality. The implementation of QFL promotes quality in the organization and the integration of software analytics and project planning tools also improves the communication between teams.*

*Keywords— Quality Management Process, Quality Requirement, Quality Assessment, Software Analytics Tool, Quality Monitoring, Requirements Pattern.*

I. INTRODUCTION

Software quality is an essential competitive factor for the success of IT companies today. Market prospects indicate that up to 26% of firms' IT budgets are dedicated to software quality assurance and testing, and they predict an increase to 33% in the next three years [1]. A report conducted by the Tricentis software testing company revealed that software failures caused more than $1.7 trillion in financial losses in 2017 [2].

Agile software development (ASD) has been adopted by organizations as a way of reducing time-to-market without hampering quality. Related to not hampering quality, Quality Requirements (QRs) are defined as "a requirement that pertains to a quality concern that is not covered by functional requirements", playing an important role in the success of software projects [3]. QR management is still a challenge in ASD contexts, *"Limited ability of ASD to handle QRs'"* is the most reported challenge in the QR management in agile [4]. Some deficiencies are reported, like lack of techniques for elicitation or linking QR to functional requirements, and user stories inadequate to specify this kind of requirements.

In the context of ASD, a large number of tools are used during the development producing a big amount of data. Software analytics tools provide features for analysing and visualizing this data to support data-driven decision-making [5][6]. The aim of this study is to explore the benefits of integrating software analytics tools to assess software quality and support QR management processes. In this context, this paper presents an experience of continuous assessment and monitoring of QR using the Q-Rapids software analytics tool (Q-Rapids Tool for brevity). Its design aligns with the guidelines defined by Buse and Zimmermann [7] such as easiness of use and interactivity. The functionalities of Q-Rapids Tool are based on the definition of product and process quality-related indicators to assess software quality. Q-Rapids Tool includes a concrete feature supporting QRs elicitation, by providing semi-automatic QR generation based on the product quality-related indicators assessment. The Softeam company has integrated Q-Rapids Tool into their quality assessment process allowing them, not only to assess product quality, but also to monitor the QR development process.

According to Ochodek' survey on the perceived importance of some agile RE practices, *"the most critical agile Requirement Engineering practices are those supporting iterative development with emergent requirements and short feedback loop"* [8]. The Quality Feedback Loop (QFL) process presented in this paper, supported by Q-Rapids Tool, has been designed with the purpose of providing decision-makers continuous feedback about the QRs emerged from their product quality assessment.

The rest of this paper is structured as follows. First, Section II includes the details of Q-Rapids Tool and Section Section III details of the methodology applied. Then, Section IV presents the Modelio case study from Softeam, describing the context of this study. Section V presents the Quality Feedback Loop (QFL) process supporting QR management, and Section VI the details about it has been implemented by Softeam, including the used tools. The benefits, challenges and lessons learned provided by the QFL users about the QFL implementation are included in Section VII. Finally, Section VIII closes the paper with the conclusions of this study.

II. Q-RAPIDS SOFTWARE ANALYTICS TOOL

This section describes the Q-Rapids software analytics tool integrated into Softeam's development process as part of this study. In order to fully understand the functionality of the selected tool, the initial subsection describes how the data is aggregated by the tool in a 3-layer quality model.

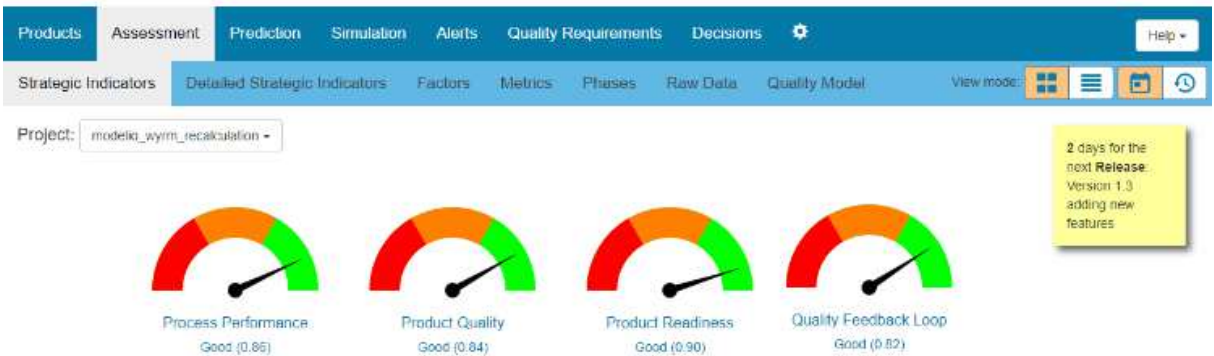

Fig. 1. Q-Rapids Tool Dashboard (visualization of Strategic Indicators)

## A. Q-Rapids Quality Model

The quality assessment provided by Q-Rapids Tool, described in the following sub-section, follows a 3-layered structure imposed by the Q-Rapids quality model [6]. The three layers define quality metrics, quality factors, and strategic indicators:

- *Quality Metrics* are low-level indicators measuring a specific characteristic (e.g., *Comments Ratio*) assessed from the raw data (e.g., *issues*) stored in data source tools (e.g., *JIRA*).

- Quality Metrics are aggregated into *Quality Factors* (e.g., *Code Quality*, *Tasks' Velocity*) that define significant concepts in relation to quality.

- Quality factors are combined to compute the higher-level model elements named *Strategic Indicators* (e.g., *Product Quality, Process Performance*). Strategic indicators allow representing the level of achievement of those aspects that companies consider relevant to their software products and decision-making processes.

TABLE I. includes the definition of the two quality factors used by Softeam for the assessment of the *Product Quality* strategic indicator. Detailed Softeam quality model included in Section IV.B.

TABLE I. PRODUCT QUALITY METRICS

| *Code Quality* quality factor definition | | |
|---|---|---|
| **QM Name** | **Definition** | **Data Source** |
| Comments Ratio | Percentage of files laying within a defined range of comment density | SonarQube analysis tool |
| Complexity | Percentage of files not exceeding a defined average cyclomatic complexity (based on the number of paths through the code) | SonarQube analysis tool |
| Duplication Density | Percentage of files laying within a defined range of duplicated blocks of lines of code | SonarQube analysis tool |
| *Critical Issues* quality factor definition | | |
| **QM Name** | **Definition** | **Data Source** |
| Specification Task Completeness | Percentage of specification tasks defining quality criteria that have been completed | Project backlog tool |

## B. Q-Rapids Software Analytics Tool

Q-Rapids Tool has been developed as a result of the Q-Rapids European research project (www.q-rapids.eu). Q-Rapids Tool's main objective is to provide continuous quality assessment to support decision-makers in the management of quality in agile software development. Besides, the tool provides concrete functionalities to support decision-making processes related to QR management. In the following, we describe its main functionalities [9].

*Quality Assessment*. Q-Rapids Tool gathers data from heterogeneous external data sources, such as static code analysis (e.g., SonarQube), continuous integration tools (e.g., Jenkins), code repositories (e.g., SVN, Git, GitLab), issue tracking tools (e.g., Redmine, GitLab, JIRA, Mantis), and usage logs. This data is aggregated into strategic indicators following the quality model described in the previous sub-section, the data is presented to decision-makers through a dashboard (see Fig. 1).

*Predictions and What-if Analysis*. These features assist decision-makers in their decisions related to the product evolution. A number of available prediction techniques (e.g., ARIMA, ETS, Neural Networks) forecast the evolution of the strategic indicators over time, considering the trends of the data sources [10][11]. This functionality gives the opportunity to react to potential issues related to concrete aspects of the product. When prediction points out possible upcoming quality issues, what-if analysis can be used to assess improvement options showing how these changes would impact on the strategic indicators, i.e., which would be the SI assessment responding to concrete factors or metrics values. In fact, what-if analysis can be used as an independent at any moment during the quality assessment process.

*Quality Requirement Semi-automatic Generation*. When the quality assessment is below some user-defined thresholds for one or more quality model elements, a quality alert is generated. Q-Rapids Tool suggests candidate QRs related to these alerts and visualizes the impact on the strategic indicators of implementing the suggested QR. In order to suggest QR, Q-Rapids Tool includes a QR pattern catalogue that should be refined by the organization to address the quality aspects included in their quality model [12].

## III. RESEARCH APPROACH

The goal of this study is to explore the benefits of integrating the Q-Rapids software analytics tool to assess software quality and support QR management processes. In order to achieve this goal, we formulated the following research questions:

- *RQ1. Can we integrate QR management into Softeam's development process?*
- *RQ2. Can we integrate software analytics tool Q-Rapids in the Softeam's toolchain?*
- *RQ3. Can we verify the effectiveness of the QR management process?*

To answer these questions, we followed a technology transfer approach distinguishing two phases: formative and summative. The formative phase's goal was understanding how to integrate Q-Rapids Tool in Softeam's development process and toolchain. This process was conducted in a series of workshops where the participants discussed concrete development activities that can benefit the use of Q-Rapids Tool. These workshops were conducted in the period from February to May 2019. Participants included company members playing quality manager, project manager, and developer roles. As a result, we defined incrementally the Quality Feedback Loop (QFL) process supported by the integration of Q-Rapids Tool and Open Project (Softeam's planning tool), answering RQ1 and RQ2. In order to integrate Q-Rapids Tool in QFL, we specialised the baseline QR catalogue integrated into Q-Rapids Tool, to fit the quality expectations for Modelio, and identified the thresholds that should raise the alerts expected in the process. As part of QFL, the quality manager and the project manager suggested the definition of a dedicated strategic indicator in Q-Rapids Tool to verify the effectiveness of the process. The indicator was named Quality Feedback Loop and its definition allows answering RQ3.

The summative phase was conducted in the period from May to June 2019. In this phase, Q-Rapids Tool was used in two pilot projects (Modelio NG and Modelio Wyrm). Some Modelio team members (project owner, project manager, quality manager, developer, and DevOps engineer) submitted monthly reports informing on the use of the tool. A final summative evaluation session was conducted in June 2019. During a one-day workshop, the participants discussed the application of Q-Rapids Tool and QFL under realistic circumstances, the related benefits, and the strengths of the Q-Rapids Tool. Part of the workshop results are reported as lessons learned in this paper.

## IV. THE MODELIO CASE STUDY

Softeam[1] is a French consulting and technology service company with over 30 years of experience in building modelling environments, over 1400 employees and operations in London, Singapore and Paris. One of their products is the open source Modelio [2] modelling environment software product line. Modelio allows modellers to create and manage models in various notations, including UML, BPMN, SysML, TOGAF and SoaML, among others. It is licensed under the GNU General Public License version 3 and it is written in Java, being implemented as an Eclipse Rich-Client Platform application. Modelio includes its own scripting environment for various modelling tasks (e.g., code generation), based on the Jython implementation of the Python programming language.

The experience reported in this paper has been conducted in the context of Modelio development. The development team applies a Scrum-based development process providing a release every six months. Q-Rapids Tool was used to continuously manage the quality of the software artefacts produced and monitor the development process (details on the use included in Section VI.A). Q-Rapids Tool has been used from September 2018, in three released versions: Valkyrie (December 2018), NG (April 2019), and Wyrm (October 2019).

### A. Softeam's Development Process

The Softeam development process (Fig. 2) starts with an initial specification phase aiming to identify the main features that will be included in the next release and solve key architectural questions resulting from previous releases (activity *Defining Features*). This initial phase is followed by the development phase, implemented as iterative development sprints decomposed each into three main activities: (i) Plan Feature (*Whiteboard Meeting*) aiming to define the tasks that which will be implemented during the next sprint, (ii) *Sprint Implementation* which covers all development tasks, and (iii) *Sprint Evaluation* which aims to assess the output of the sprint in order to update the project plan and prepare the next sprint iteration. The results of the sprint evaluation can lead to some changes in the process (*Process Adaptation* activity). In the pre-release validation phase, the *Integration and Validation* activity includes a final quality assessment of the product to ensure the quality of what will be delivered to customers.

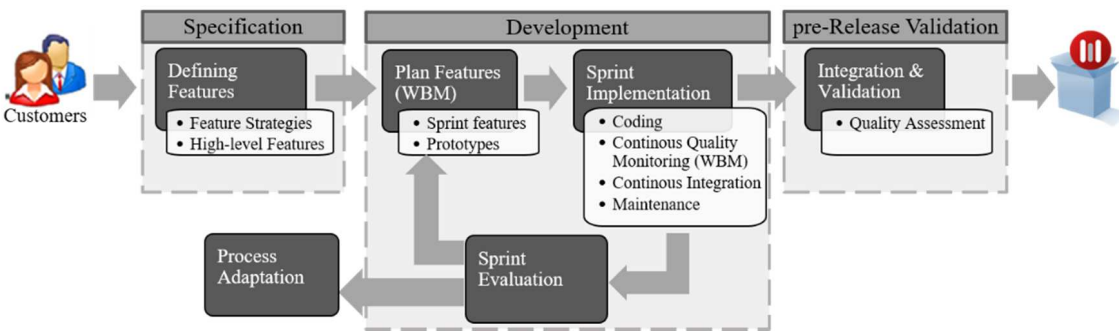

Fig. 2. Softeam development process

---

[1] https://www.softeamgroup.fr/en/   [2] https://www.modelio.org/

The *Sprint Evaluation* is conducted by *Quality Engineers*, who are responsible for the specification of the general quality criteria and the definition and implementation of the quality controls performed during the development process. They also monitor the development process to identify potential problems and areas for improvement. The *Project Manager* oversees all the work required for the on-time delivery of the product, including the expected features and desired level of quality. Regarding quality management, the project manager is responsible for identifying and planning tasks aimed at resolving the quality problems reported by the quality engineer. If it appears that the development process must be adapted, the Project manager is also responsible for implementing the necessary changes.

*B. Softeam Quality Assessment*

As part of the customisation of Q-Rapids software analytics tool, Modelio team defined three strategic indicators for assessing quality of Modelio product and development process: *Process Performance*, *Product Quality*, and *Product Readiness* (see Fig. 3). In order to compute these three strategic indicators, there needed nine quality factors and twenty-two quality metrics. Data needed for the quality metrics computation is gathered from six data sources (Mantis, Open Project, SonarQube, Jenkins, SVN, and Modelio logs).

*Process performance* refers to software development lifecycle processes' efficiency and quality; *Product Quality* refers mainly to internal quality; and, *Product Readiness* provides information used to know if the product is ready to be released, i.e., implements the scheduled features with no critical issues open.

## V. QUALITY FEEDBACK LOOP PROCESS

In agile development contexts, a big amount of data is produced during development due to the high number of tools used in the different development phases (e.g., issue track systems, source code versioning, testing, integration...). Software analytics tools use this data to help decision-makers provide evidence to support their decisions. In the context of Requirements Engineering, the use of these tools is twofold: (i) validating the QRs included in the product backlog, and (ii) making evident some code quality shortcoming, which can be mitigated by proposing new QRs.

We integrate both aspects in the definition of the Quality Feedback Loop (QFL) process. The QFL process covers the whole life-cycle of QR management, from QR elicitation to their validation, integrating the use of software analytics tools. QRs are critical to produce a successful product, but they are not the only factor to succeed. Another important factor to consider is the development process quality. Therefore, QFL also controls the QR development process as part of the product quality control. Fig. 4 presents the QFL cyclic process that is composed of three phases: *QR elicitation*, *QR planning* (i.e., QR integration in the project plan), and *QR feedback monitoring* closing the cycle, which are detailed in the following subsections

*A. Quality Requirements Elicitation*

In agile software development, QRs can come explicitly from the customers or architects or as an outcome of the quality assessment of the product. Softeam complements the explicit QRs elicited upfront, with QRs generated semi-automatically by Q-Rapids Tool, based on the quality assessment of their products.

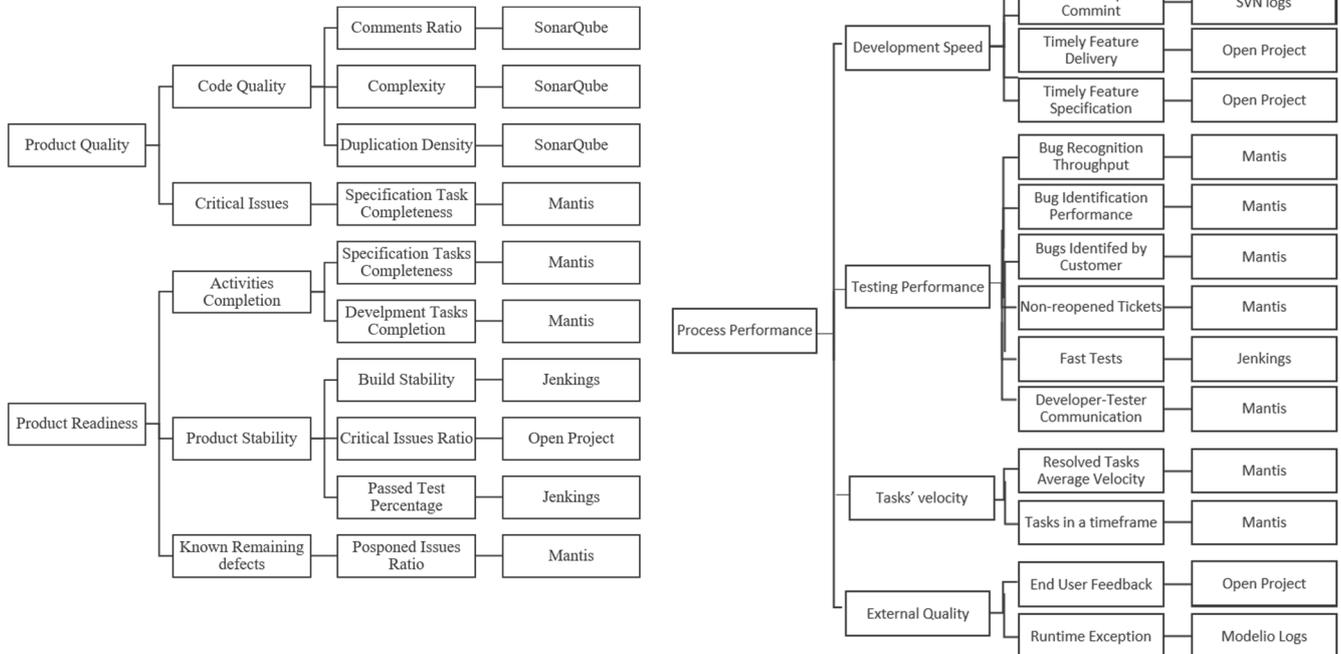

Fig. 3. Softeam's Quality Model implented in Q-Rapids software analytics Tool

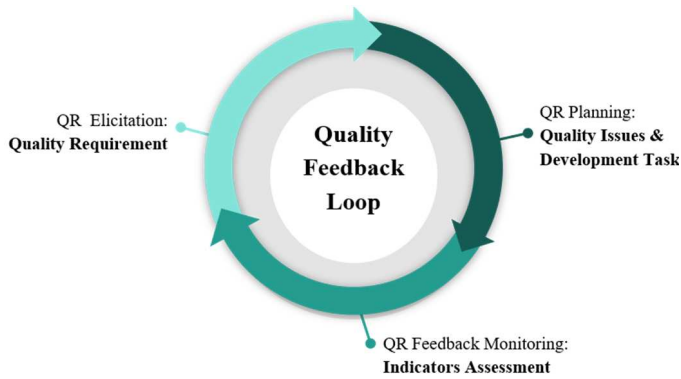

Fig. 4. Quality Feedback Loop process

In order to support semi-automatic QR elicitation, Q-Rapids Tool uses a QR pattern catalogue [12]. The Softeam QR pattern catalogue contains six patterns organised into seven categories (some patterns are classified into more than one category). QR patterns are defined to address concrete qualities included in the quality model (i.e., quality metrics) and categories correspond to quality factors (see Table 1). The connection between the patterns and the quality model elements allows the tool to identify candidate QRs when the assessment of the quality model elements reveals some shortcoming. For example, the catalogue includes three patterns related to quality factor *Code Quality* (see Table 1 for factor definition): one that can be used when the *Comments Ratio* quality metric is not good enough, one for Complexity quality metric, and one for *Duplication Density* quality metric. TABLE II. shows the information of the *Complex Files* pattern, linked to the *Complexity* quality metric, which should be instantiated by the decision-maker in case it is selected as a candidate to be included in the product backlog.

In addition, this catalogue could be eventually used by decision-makers to elicit QRs manually using the knowledge gathered from other projects that are feeding the catalogue.

The fact of using a quality model for the quality assessment provides a suitable scenario for the use of patterns. The definition of software quality is systematized in a way that the desired organization system qualities are analysed and identified making easier the maintenance of the desired QRs in a form of patterns, being instantiated depending on the project they are going to be included.

TABLE II. COMPLEX FILES PATTERN

| Name | Complex files |
|---|---|
| Description | Ratio of non-complex files (defined as files with a cyclomatic complexity below 15) with respect to the total number of files should be at least the given value in order to improve the quality of the source code. |
| Goal | Improve the quality of the source code |
| Pattern text | Ratio of non-complex files should be at least `%value%` |
| Parameter description (between `%%` in the pattern text) | |
| Name | value |
| Correctness | 0 <= value <= 100 |
| Description | value in percentage of the maximum percentage that acceptable complex files in relation of the ratio of open/in progress |
| QM metric | complexity |

### B. Quality Requirements Planning

Elicited QRs need to be integrated in the product backlog. In Softeam, they took the form of Quality Issues, i.e., backlog items devoted to document QRs. When a quality issue is moved to the development phase, it is decomposed into concrete development tasks in the sprint backlog, similarly to functional-oriented user stories. The introduction of a specific type of backlog items for the QRs meets several expectations.

Firstly, having a dedicated issue type in the backlog allows decision-makers to monitor the implementation of the associated development tasks independently of other project activities and thus produce specific indicators to assess how quality issues are addressed in this development process.

Secondly, decision-makers are reluctant to the idea that tools can automatically modify the project plan by generating product backlog items. On the other hand, it is necessary to ensure that all detected quality issues are addressed and keep track of the actions taken to solve the problem by them. Managing QRs as high-level issues is a way to leave the final decision to the decision-makers. Including all the generated QR in the backlog is ensuring the traceability of quality requirements to the development tasks. A quality issue can be rejected by the decision-maker, postponed or refined as concrete action.

### C. Quality Requirements Feedback Monitoring

In the QR feedback monitoring phase, the decision-maker analyses feedback from two perspectives: (i) assessing the quality of the resulting software, thus validating the elicited QR; and (ii) controlling the QR development process.

The quality model, used for the quality assessment in the tool, defines the desired system qualities in a measurable form (i.e., quality metrics), providing the means for the QR validation. The quality model elements are linked to the QR in the backlog. This traceability provides control on the QR development process, allowing decision-makers to control two process aspects: the effectiveness of the automatic generation of candidate QR generation and the progress of the development tasks associated with these QRs. To integrate the process control, Softeam defined a strategic indicator in the Q-Rapids Tool named *Quality Feedback Loop* that measures both aspects (see TABLE III. ). It aggregates two factors: *Quality Requirements Relevance* and *Quality Requirements Completion*.

TABLE III. FEEDBACK QUALITY LOOP METRICS

| *Quality Requirements Relevance* quality factor definition | |
|---|---|
| QM Name | Definition |
| Quality Requirements Acceptance | Percentage of quality requirements not rejected by project manager |
| *Quality Requirements Completion* quality factor definition | |
| QM Name | Definition |
| Mitigation Task Completion | Percentage of development tasks derived from quality requirements which has been completed |
| Quality Requirement Derivation | Percentage of accepted quality requirements derived as concrete development tasks |

The *Quality Requirements Relevance* factor measures whether the QRs identified in the elicitation phase (quality issues) are pertinent for the project (accepted for implementation). The *Quality Requirements Completion* factor measures the progress of the accepted QRs that are in the development phase (i.e., derived into development tasks). The data source for all the metrics is the Project backlog tool.

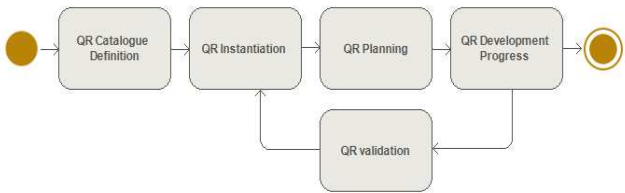

Fig. 5.  QR management life-cycle

## VI. QUALITY FEEDBACK LOOP IN USE

In this section, we describe how the QFL process has been implemented in Softeam. This implementation is described in terms of how Softeam has integrated QFL and its associated tools (software analytics and project planning) in their development process and how these tools have been used in the QFL phases.

Fig. 5 includes the activities covered by the QFL process, which covers the QR management life-cycle as follows: QR elicitation, including QR catalogue definition and QR instantiation in context of a project; QR planning; and, QR development and QR validation, as part of the QR feedback monitoring.

### A. QFL Integrated into the Softeam Development Process

In the context of the Softeam development process (see Fig. 2), QFL is affecting the development phase (*Plan Features*, *Sprint Implementation*, and *Sprint Evaluation* activities) and the pre-release validation phase (*Integration & Validation* activity). Fig. 6 depicts the Softeam software development process integrating the QFL process and the tools supporting each phase (Q-Rapids Tool and Open Project). The figure also includes the data source tools, used by Softeam, producing data consumed by Q-Rapids Tool for assessing quality.

During the *Sprint Implementation* activity, development data is generated from several tools (Redmine, Mantis, OpenProject, SVN, Mantis, Jenkins, and Modelio Testing System), and it is collected and analysed by Q-Rapids Tool to assess product and process quality. The data is used to compute quality metrics, quality factors and strategic indicators on the bases of Softeam's quality model.

Quality metrics assessment (QR feedback monitoring) is used by *quality engineers* and *project managers* in the *Sprint Evaluation* activity to identify quality issues in delivered artefacts (QR elicitation), which are exported from Q-Rapids Tool to OpenProject tool. The development process is also analysed, taking special attention to activities addressing quality problems.

The team uses analysis reports produced in the sprint evaluation, in conjunction with the quality issues already included in the OpenProject, to plan the next sprint iteration (*Plan Feature activity*). The *project manager* decides to accept (or not) the quality issues, and for the accepted ones, the best way to implement them by transforming them into development tasks included in the upcoming sprint backlog. In the *Integration & Validation* activity, Q-Rapids Tool is also used to check the alignment of the release candidate with quality constraints defined by Softeam for each now product provided to its customers. When this analysis reveals some quality shortcoming, the project manager can decide not to include some features or postpone the release.

During the sprint evaluation, if quality metrics and strategic indicators related to the development process highlight major problems, the process can be adapted for the next sprint (*Process Adaptation* activity). These adaptations can take various forms, such as the revision of the quality criteria applied to all of the products (adjusting the quality model), the adjustment of the threshold values from which an alert is triggered, or the updating of the QR catalogue by introducing new patterns or by modifying the criteria for applying existing ones.

### B. QFL for QR Elicitation and QR Planning

Q-Rapids Tool provides a new mechanism that allows Softeam to control whether a product meets their quality criteria; this feature is the semi-automatic QR generation. Fig. 7 depicts the tools interoperability.

The setting up of the QR generation system started by the configuration of alerts' thresholds (qr-alert component; step 1 in Fig. 7). Q-Rapids Tool periodically checks the quality metrics computed by the tool according to customisable thresholds defined as quality goals by Softeam Quality Engineers. If a monitored quality metric exceeds the identified thresholds, the Softeam Quality Team is notified via Q-Rapids Tool's dashboard (qr-dashboard; step 2). An adapted answer, in the form of parameterizable QR (from the QR pattern catalogue), is suggested to the Quality Engineers who can choose to include it in the backlog. In this case, a QR candidate presenting a solution to address the quality issues will automatically be generated (qr-backlog; step 3) in Softeam's project management tool, Open Project.

.

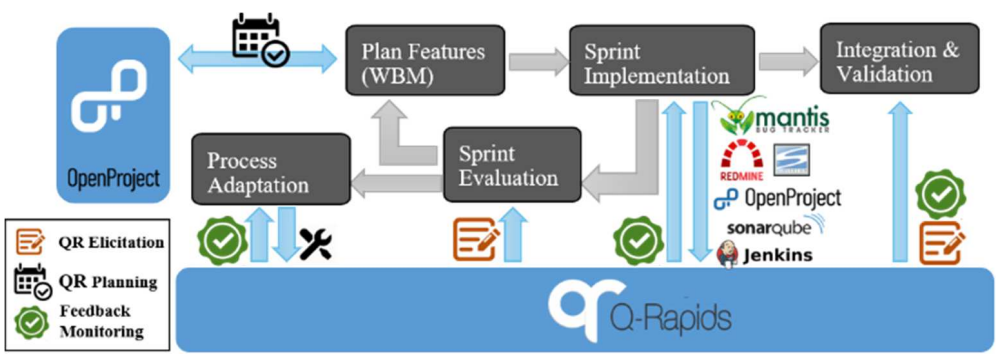

Fig. 6.  Quality Feedback Loop integrated in Softeam' software development process

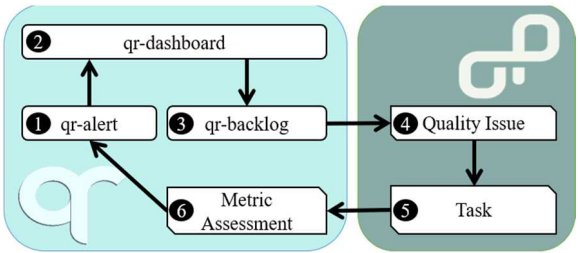

Fig. 7.  QFL tool chain in Softeam

On the OpenProject side, the generated QR candidate is integrated into the project roadmap as a quality issue. Quality issues are implemented as a new type of abstract Work Package task named "*QualityRequirement*". Softeam made the decision to not directly generate concrete tasks in the project roadmap, allowing the project manager to choose the best approach to solve the identified quality issue. The project manager decides to keep or reject the quality issues. For the accepted quality issues, the project manager will derive development tasks when he or she decides to move it to the development phase

Fig. 8 includes examples of QRs generated in Softeam's OpenProject tool. Some of them accepted and deriving one or more development tasks (e.g., 133 and 134), and some of them rejected (e.g., 135).

To evaluate the usability of the semi-automatic QR generation, we requested quality engineers and project managers monthly reports defining usage stories, i.e., definition of real scenarios where they use this feature and the purpose of the usage following the next template:

> As a <role>, I have used <tool/tool feature>, to <purpose>, during <activity>.

Following, there is an example of usage scenario reported by a project manager: "*As a Project Manager, I have used OpenProject to receive notification related to a potential quality issue in Modelio NG by the intermediary of new quality requirements generated into my Project Management Tool (related to Ratio of open issues, and the ratio of properly commented files), during the Whiteboard Meeting*". The Whiteboard meeting corresponds to the plan features (WBM) activity.

Complementing the scenario usage, the user could also report about the strengths and the general satisfaction. The same user reported the level of satisfaction as "Happy" and the following strength "*I was able to decide to include a new task in project development platform to address the 'Ratio of open issues' QR and to reject the 'ratio of properly commented' (this issue will be addressed by a reminder to developer about coding rules)*".

The Softeam's developed QRs Generator Open Project Connector is publicly available at GitHub[3]. We developed a specific Spring Boot based extendable REST API to transform QRs generated by Q-Rapids Tool into OpenProject work items. Information on backlog Services specification is at GitHub[4].

*C. QFL for QR feedback monitoring*

Q-Rapids Tool provides quality assessment visualization that can be used for QR monitoring and validation, and for QR development process assessment.

Related to QR monitoring and validation, the tool includes an historical view with information about the decision of adding a QR. Fig. 9 (left) shows the adding QR decisions as green crosses. This visualisation allows us to analyse if the QR development matches the expectations when the QR was elicited, i.e., the assessment of the related quality model element increases after the decision. The chart shows how *Passed Tests Percentage* metric assessment improves 20% after adding the QR "*The percentage of passed automatic tests should be at least 0.95*" on June 26th.

| ID | SUBJECT | TYPE | STATUS |
|---|---|---|---|
| 117 | Modelio SAS Support | Feature | New |
| 118 | Archimate 3.0.1 | Feature | Closed |
| 119 | Constellation for Modelio SAS | Feature | Closed |
| 133 | Ratio of open issues of type bug should be below 5% | QualityRequirement | Closed |
| 136 | Initiate a bug fix session adressing Archimate Issues | Task | Closed |
| 134 | Ratio of files without critical or blocker quality rule violations should be below … | QualityRequirement | Closed |
| 138 | Review the validation process of Modelio components | Task | Closed |
| 156 | Refactor and evaluate validation process | Task | In progress |
| 135 | Ratio of open issues of type bug should be below %value% | QualityRequirement | Rejected |

Fig. 8.  Quality Requirements in Softeam's OpenProject Tool

---

[3] https://github.com/q-rapids/qrapids-backlog-openproject

[4] https://github.com/q-rapids/qrapids-dashboard/wiki/qrapids-backlog-Services

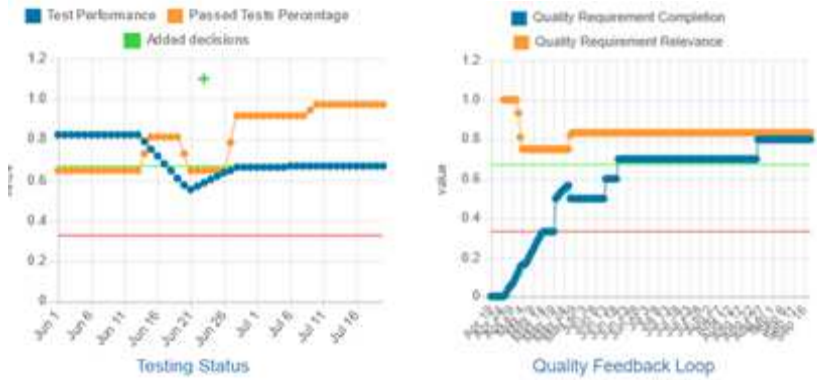

Fig. 9. Testing Status (left) and Quality Feedback Loop (right) in historical views

Related to QR development process, using the strategic indicator *Quality Feedback Loop* (TABLE II. ), the project manager can control the evolution of the relevance of the elicited QR (Fig. 9, right, orange series) and, for the QRs that he or she has accepted, the process of closing the development tasks associated with them (Fig. 9, right, blue series).

*Quality Requirements Relevance* factor assesses the quality of the candidate QRs generated by the tool. Low values in this indicator are alerting about bad configuration of the alerting component (i.e., not suitable thresholds). *Quality Requirement Completion* factor is assessing the development progress of the generated QRs that are already in the development phase (i.e., with derived development tasks in the backlog).

Fig. 9 (right) shows how the *Quality Requirements Relevance* factor has some fluctuations at the beginning of the period, arriving at an stable assessment after the tuning of the thresholds that trigger the alerts. *Quality Requirement Completion* factor shows how the development has been completed during the period, following the expected behaviour, being low at the beginning and high at the end.

The quality assessment visualization, complemented by some other features provided by the tool (prediction and what-if analysis), also supports the quality engineer and project manager in the task of manually identifying new QRs, closing the feedback quality loop.

During the period from May to December 2019, seven QRs were suggested by Q-Rapids Tool based on quality alerts. Six of them were accepted by quality engineers. From these six, four were confirmed by project managers, being translated to development tasks, and two rejected. As a result, 85% of the QR suggested by Q-Rapids were accepted by quality engineers, from them 66% were accepted by project managers. In other words, 57% of the QR suggested by Q-Rapids Tool ended in the product backlog included in the product roadmap, with plans of being developed by the development team.

## VII. CHALLENGES AND LESSONS LEARNED

As a result of the implementation of QFL, Softeam has automated their quality management process by identifying quality issues, using the alert mechanism, and including quality requirements into Softeam's product backlog. QFL also provides them the monitorisation of their quality requirements resolution process. They can check which identified quality issues have been addressed by their development cycle and follow the progress of the implementation of mitigation actions dealing with quality issues.

The QFL users identified several challenges in the integration of a software analytics tool and the QFL process [13]. First, the effort required to deploy a software analytics platform (i.e., Q-Rapids) in new projects is considerable. User interfaces to configure data gathering components, avoiding technical knowledge, could be important contributions. Second, it is necessary to dedicate efforts to work on the integration of the software analytics platform in the organization existing development process, which is only possible when the adaptation to the practices of the organization is compliant. A third challenge is building the confidence of operational teams in the metrics and indicators to be implemented by the software analytics platform. And fourth, there are legal and privacy impediments to collect user usage data (e.g., logs from our customer's clients) which deserve attention from the research community.

Below, we summarise the lessons learned provided by the Modelio team as a summary of the summative evaluation of QFL.

*Tool interoperability*. The QR integration approach in the OpenProject tool was very effective allowing our Softeam's project managers to completely control the whole process and to make the final decisions.

*Semi-automatic QR generation*. Although quality is essential for Softeam's teams, none of Softeam's project managers was willing to accept the project plan to be directly modified by a supporting tool without supervision. To solve this issue, we decided that QRs to be integrated into the project work package as a new kind of task type named "*QualityRequirement*". From these abstract tasks, we left the project manager the possibility to decide to close the task if it considered that the recommendation is not relevant, or to derive the abstract task into a concrete action integrated in the project roadmap if it is relevant. In both cases, the project manager must justify the decision by providing a rationale. The project manager's response to these quality issues is further monitored by Q-Rapids tool in order to produce indicators on the quality process implementation. *Promoting quality into the organization*. This new approach of managing quality in Softeam projects (QFL process) has stirred up a lot of enthusiasm both from Softeam's quality engineers and project managers. The quality teams see a way to promote quality in the organization, while project managers appreciate the recommendation integrated to project management tools and freedom that have left them on the manner of dealing with these problems. All project stakeholders have also highlighted the increasing benefits in communication between the various teams allowed by the automation of the QR management process enabled by Q-Rapids Tool.

*Knowledge reuse across the organization*. Completion of the QR catalogue during projects allowed a formalization of the way in which the quality problems are addressed in software development processes at Softeam. This constant improvement is enabled by the monitoring mechanisms of Softeam's quality management processes embodied by the *Quality Feedback Loop* strategic indicator. Most QRs not being specific to a particular project, distributing this catalogue to the various development teams of the company helped disseminate good practices relating to quality management and significantly improves the quality of the products delivered by Softeam.

## VIII. CONCLUSIONS

In this paper, we present the experience of Softeam, a French consulting and technology service company, using a software analytics tool (Q-Rapids Tool) for continuous assessment and monitoring of Quality Requirements (QR). The integration of this tool into their quality assessment process resulted in the definition of the Quality Feedback Loop (QFL). QFL covers the QR management life-cycle in the context of agile software development as follows: QR elicitation, QR planning, and QR feedback monitoring. QFL process allowed us to integrate QR management into Softeam's process (*RQ1*).

Q-Rapids Tool has been successfully integrated into Softeam's toolchain (*RQ2*). This integration includes the data gathering needed from Q-Rapids Tool to assess Modelio's quality (Mantis, Open Project, SonarQube, Jenkins, SVN, and Modelio logs) and the integration with the Softeam's project management tool Open Project to implement QFL. Q-Rapids Tool uses product quality assessment and a QR pattern catalogue to identify candidate QR that are suggested to the quality engineers. Quality engineers can decide to include them in the product backlog, and then these QRs are exported to the OpenProject tool. The project manager needs to accept (or reject) the QR included in the product backlog (Open Project) by quality engineers. For the accepted ones, project managers derive the corresponding development tasks in OpenProject to be implemented by the development team.

Q-Rapids Tool provides mechanisms to monitor the quality of the product defining indicators devoted to measure the desired qualities (QRs validation). Softeam defined the strategic indicator *Quality Feedback Loop,* monitoring the QRs development progress of the corresponding development tasks, to verify the effectiveness of the QR management process (*RQ3*).

From the experience of using QFL from May 2019 to December 2019, four of the seven (57%) QR suggested from Q-Rapids Tool ended in the product roadmap, with plans of being developed by the development team.

Referring to the concrete software components, the detection of anomalies is performed by Q-Rapids Tool (qr-alert component), based on quality assessment, and the generation of quality requirement is triggered manually by quality engineer using Q-Rapids Tool dashboard (qr-dashboard web application). In order to integrate Q-Rapids Tool and OpenProject, Softeam had developed the qr-issuetracker-openproject plugin allowing to generate quality requirements into the end user development process tool (OpenProject).

The users of QFL highlighted as strengths the integration of Q-Rapids Tool QR generation with their project management tool (OpenReq) and the fact of giving the final word to quality engineers and project managers in the decision of including (or not) the QR in the product backlog. As a collateral consequence of the QFL definition and implementation, the quality teams see a way to promote quality in the organization. The integration of both tools also improves the communication between teams.

### ACKNOWLEDGMENT

This work was supported by Q-Rapids (Quality-Aware Rapid Software Development. Q-Rapids was funded by the European Union's Horizon 2020 research and innovation programme under grant agreement nº 732253. We thank all members of Softeam who participated in the evaluation of the quality feedback loop proposed in this paper.